# Summary Statistics from Training Images as Prior Information in Probabilistic Inversion

Tobias Lochbühler[1,2], Jasper A. Vrugt[2,3], Mojtaba Sadegh[2], Niklas Linde[1*]

[1] Applied and Environmental Geophysics Group, Institute of Earth Sciences, University of Lausanne, Switzerland;

[2] Department of Civil and Environmental Engineering, University of California Irvine, Irvine, CA, USA;

[3] Department of Earth System Science, University of California Irvine, Irvine, CA, USA.

*corresponding author: Niklas.Linde@unil.ch

**Abstract**

A strategy is presented to incorporate prior information from conceptual geological models in probabilistic inversion of geophysical data. The conceptual geological models are represented by multiple-point statistics training images featuring the expected lithological units and structural patterns. Information from an ensemble of training image realizations is used in two different ways. First, dominant modes are identified by analysis of the frequency content in the realizations, which drastically reduces the model parameter space in the frequency-amplitude domain. Second, the distributions of global, summary metrics (e.g., model roughness) are used to formulate a prior probability density function (pdf). The inverse problem is formulated in a Bayesian framework and the posterior pdf is sampled using Markov chain Monte Carlo simulation. The usefulness and applicability of this method is demonstrated on two case studies in which synthetic crosshole ground-penetrating radar travel time data are inverted to recover 2-D porosity fields. The use of prior information from training images significantly enhances the reliability of the posterior models by removing inversion artifacts and improving individual parameter estimates. The proposed methodology reduces the ambiguity inherent in the inversion of high-dimensional parameter spaces, accommodates a wide range of summary statistics and geophysical forward problems.

# 1    Introduction

Geophysical inversion seeks to infer representative estimates of spatially distributed subsurface properties, given direct or indirect measurements that are sensitive to those properties. For a specified model parameterization, each set of parameter values is referred to as a model. In probabilistic inversions, the probability of a certain model given the data is expressed as the product of a prior distribution and a likelihood function using Bayes' Theorem. The prior distribution is a probability density function (pdf) that summarizes all information available about the parameters of interest before any data is collected, whereas the likelihood function quantifies the probability that the actual data with its measurement errors has been generated by a proposed model. The product of the prior distribution and likelihood function leads to a posterior distribution, which summarizes the statistical distribution of the model parameters. Unfortunately, in most practical applications we cannot derive this posterior distribution by analytical means. We therefore resort to Monte Carlo simulation to generate samples from the posterior distribution. The most popular of such sampling methods is the Random Walk Metropolis (RWM) algorithm (e.g., Chen et al., 2000), a Markov chain Monte Carlo (MCMC) simulation method that has found widespread application and use in many different fields of study. Such MCMC methods have several advantages over deterministic, gradient-driven optimization methods in that they (1) do not require linearization of the inverse problem and thus can handle highly non-linear problems and (2) provide an ensemble of models drawn from the posterior distribution that can be used to characterize parameter uncertainty. Recent applications of MCMC in geophysics include inversions of electromagnetic (Buland and Kolbjørnsen, 2012; Rosas-Carbajal et al., 2014), seismic (Bodin and Sambridge, 2009; Hong and Sen, 2009) and ground-penetrating radar data (Scholer et al., 2012), as well as multi-physics inversions (Bosch et al., 2006; Irving and Singha, 2010). Sambridge and Mosegaard (2002) present a comprehensive but somewhat dated review on Monte Carlo methods in geophysical inverse modeling.

A general problem of parameter estimation in geophysics is that there exists an infinite number of models that can explain the data within their error bounds (Backus and Gilbert, 1970), but most of them are unrealistic or even unphysical. To overcome this ambiguity and reduce the number of possible models, it is common practice to impose data independent constraints on the model structure. Constraints can be imposed by choosing model parameterizations where the model space is bounded to exclude undesired models (e.g., Lochbühler et al., 2014a), or by favoring models based on their agreement with a reference model (i.e., damping, Marquardt, 1963), the model roughness (Constable et al., 1987; Rosas-Carbajal et al., 2014), the fit to a multi-Gaussian geostatistical model (Maurer et al., 1998; Johnson et al., 2007) or the structural similarity to another model in joint inversions (e.g., Gallardo and Meju, 2004; Linde et al., 2006). These constraints are accounted for by including penalty terms in the objective function (deterministic inversion) or by using an explicit prior pdf (probabilistic inversion). Alternatively, geostatistical inversion approaches (e.g., Irving and Singha, 2010; Mariethoz et al., 2010a; Hansen et al., 2012c) adopt a constrained prior in the form of a variogram and estimate the parameter values by sequential resimulation of parts of the model domain so that the underlying geostatistics are honored.

The geostatistical inversion approach has been extended recently to consider priors that account for higher order spatial statistics, so-called multiple-point statistics (MPS). The concept of MPS is becoming increasingly popular in the geostatistics and geophysics communities, since it allows for models that are constrained by detailed conceptual information about the expected geology at the site of interest (Caers and Zhang, 2004; González et al., 2007; Huysmans and Dassargues, 2009; Lochbühler et al., 2014b). The basis for MPS is a training image (TI), which is a graphical representation of a conceptual geological model that features the lithologies and structural patterns that could be expected at a given site (e.g., Strebelle, 2002). The MPS formulation of the spatial variability of parameter values allows for description of features that covariance-based, two-point statistics are unable to describe accurately such as curvilinear and discrete continuous structural elements. A TI thus offers strong prior information. It can be used in geostatistical inversions where MPS simulations are conditioned to geophysical data by sequential updating of groups or blocks of cells (Mariethoz et al., 2010a; Hansen et al., 2012c; Cordua et al., 2012; Hansen et al., 2012a,b). The information content of the TI drastically reduces the degrees of freedom compared with the number of cells in the model domain, and the resulting posterior models are visually very similar to the TI (e.g., Cordua et al., 2012). In these approaches, the sampling efficiency strongly depends on the number of updated cells and it can be difficult to fit large data sets with realistic observational errors (e.g., Irving and Singha, 2010; Cordua et al., 2012).

We here present an inversion strategy where TI realizations, that is, images comprising the multiple-point statistics represented by the TI, are used to define case-specific model parameterizations and prior density functions. Models are parameterized by coefficients of their truncated discrete cosine transform (DCT), with coefficients based on analysis of the TI realizations in the DCT domain (Jafarpour et al., 2009). The DCT coefficients that are necessary to describe structures that frequently occur in the TI will thus be part of the model parameterization and such structures can be reproduced in the posterior models. Apart from the model parameterization, we also extract from the TI realizations the distributions of global measures of model morphology, what we refer to as summary statistics. These distributions are translated into prior distributions to evaluate the prior probability of all proposed models.

Our inversion strategy is conceptually different from the MPS inversion approach described above, as all model parameters are sampled independently. There is also a clear distinction with respect to regularized inversions (e.g., Constable et al., 1987; Rosas-Carbajal et al., 2014; Maurer et al., 1998), in that information on the model structure is not imposed by enforcing intercellular correlation of a certain form, but instead by defining a prior on global, location-independent summary metrics. If, for example, the extracted summary metric is the global roughness, the prior probability of a proposal model depends on the global roughness distribution of the TI realizations. This distribution is unlikely to be centered around zero as imposed in classical smoothness-constrained inversions.

In the remainder of this paper, we present the individual building blocks of our inversion strategy (Section 2) and illustrate our preliminary findings for two different case studies with different geological settings (Section 3) using travel time observations from synthetic crosshole ground-penetrating radar (GPR) experiments. Section 4 discusses the advantages and limitations of the proposed methodology. Finally, in Section 5 we provide our conclusions.

# 2 Methodology

## 2.1 Sparse Model Parameterization

In applied geophysics, the model space is typically parameterized by pixels (or voxels) in 2-D (3-D) Cartesian grids. For large model domains or fine grid discretizations, this quickly results in several thousands to millions of model parameters. Since the efficiency of MCMC algorithms decreases sharply with increasing number of independently sampled parameters, inversion with a high resolution Cartesian parameterization is a daunting and CPU-intensive task, particularly in the absence of a prescribed spatial correlation structure. Fortunately, the use of sparse representations in the frequency-amplitude domain can help to significantly reduce the dimensionality of the model space while maintaining a large degree of fine-scale information. Alternative parameterizations of the model space include the discrete cosine (Jafarpour et al., 2009, 2010; Linde and Vrugt, 2013; Lochbühler et al., 2014a) and the wavelet transform (Jafarpour, 2011; Davis and Li, 2011). Here we use the discrete cosine transform (DCT, Ahmed et al., 1974), more precisely, DCT-II in 2-D, which for a uniformly discretized model $\mathbf{A} \in \Re^{N_x \times N_z}$ is given by

$$B(k_x, k_z) = \alpha_{k_x} \alpha_{k_z} \sum_{x=0}^{N_x-1} \sum_{z=0}^{N_z-1} A(x,z) \cos\frac{\pi(2x+1)k_x}{2N_x} \cos\frac{\pi(2z+1)k_z}{2N_z}, \qquad (1)$$

where

$$\alpha_{k_x} = \begin{cases} \dfrac{1}{\sqrt{N_x}}, & k_x = 0 \\[2mm] \sqrt{\dfrac{2}{N_x}}, & 1 \leq k_x \leq N_x - 1, \end{cases}$$

and

$$\alpha_{k_z} = \begin{cases} \dfrac{1}{\sqrt{N_z}}, & k_z = 0 \\[2mm] \sqrt{\dfrac{2}{N_z}}, & 1 \leq k_z \leq N_z - 1. \end{cases}$$

The matrix **B** contains the DCT coefficients. In the absence of prior information about the information content of each individual DCT coefficient, model compression can be adopted by truncating the high-frequency terms, and thus focusing on the low-frequency coefficients only (Jafarpour et al., 2009; Linde and Vrugt, 2013). In this study, we use prior information from TIs to determine case-specific sparse model parameterizations, which we hereafter conveniently refer to as TI-based parameterizations. Following Jafarpour et al. (2009), the dominant transform coefficients (i.e., those coefficients that contribute the most to represent the expected subsurface structures) are determined by the following steps:

1. create a large set of realizations of a known training image using a multiple-point statistics simulator (so-called TI realizations);
2. calculate the 2-D DCT of all TI realizations;
3. compute the arithmetic mean of the absolute value of each transform coefficient;
4. select the $n$ coefficients for which the mean value is largest, where $n$ is a default value set by the user;
5. store the coordinates in the DCT space of the $n$ dominant coefficients.

Jafarpour et al. (2009) used the reciprocals of the $n$ dominant DCT coefficients to build a weighting matrix **W** that was used as a regularization term in a deterministic inversion. In the present study, the dominant transform coefficients are the model parameters that we seek to estimate, while all other coefficients are set to zero. The DCT was preferred above the wavelet transform, as for the investigated cases the DCT had superior compression power to achieve comparable model quality.

The choice of the number of considered DCT coefficients, $n$, is a trade-off between the desired resolution and the available computational budget. The quality of the compression can be quantified by the peak signal-to-noise ratio (PSNR)

$$\text{PSNR} = 10\log_{10}\left(\frac{\max(\mathbf{A})^2}{\frac{1}{N_x N_z}\sum_{x=1}^{N_x}\sum_{z=1}^{N_z}[A(x,z)-A'(x,z)]^2}\right), \tag{2}$$

where $\mathbf{A}$ and $\mathbf{A}'$ are the uncompressed and compressed images, respectively. The PSNR is a common measure in image processing (e.g., Huynh-Thu and Ghanbari, 2008). For the two cases presented herein, we use $n = 100$ and $n = 150$, respectively, resulting in PSNR values between 28 and 39 dB. This is well above common PSNR thresholds in image processing (e.g., Welstead, 1999).

### 2.2 Markov chain Monte Carlo Sampling with DREAM$_{(ZS)}$

We seek to infer the posterior distribution of a set of model parameters $\mathbf{m}$ from observed data $\mathbf{d}$. In a probabilistic sense, the relation between the model and the data can be expressed by Bayes' Law

$$p(\mathbf{m}|\mathbf{d}) = \frac{p(\mathbf{m})p(\mathbf{d}|\mathbf{m})}{p(\mathbf{d})} = \frac{p(\mathbf{m})L(\mathbf{m}|\mathbf{d})}{p(\mathbf{d})}, \tag{3}$$

where $p(\mathbf{m}|\mathbf{d})$, the posterior pdf of the model given the data, is our main subject of interest. The other entities in eq. (3) are $p(\mathbf{m})$, the prior distribution of the model parameters, $L(\mathbf{m}|\mathbf{d}) \equiv p(\mathbf{d}|\mathbf{m})$, the likelihood of the parameters given the data and $p(\mathbf{d})$, the evidence. The latter term is generally difficult to estimate in practice, but if the model parameterization is fixed, all relevant statistical information (mean, standard deviation, etc.) can be extracted from the unnormalized distribution

$$p(\mathbf{m}|\mathbf{d}) \propto p(\mathbf{m})L(\mathbf{m}|\mathbf{d}). \tag{4}$$

In this study, the data $\mathbf{d}$ constitute the radar travel times from crosshole GPR experiments. The model vector $\mathbf{m}$ includes the dominant DCT coefficients for a given log-porosity field $\varphi$, $\mathbf{m} = \tau_{DCT} \circ \log(\varphi)$ and a set of petrophysical parameters that are necessary to solve the forward problem.

The prior $p(\mathbf{m})$ describes the distribution of the model parameters before any data **d** have been collected and assimilated (e.g., Chen et al., 2000). The formulation of the prior distribution based on summary statistics inferred from TI realizations is a central theme of this study and will be described in detail later (section 2.3).

The likelihood function $L(\mathbf{m}|\mathbf{d})$ measures the distance between the measured data, **d**, and the corresponding values predicted by the proposed model, $\mathbf{d}^{\text{pred}}$. If we assume the error residuals to be uncorrelated and Gaussian distributed, the likelihood function, $L(\mathbf{m}|\mathbf{d})$, is given by

$$L(\mathbf{m}|\mathbf{d}) = \prod_{i=1}^{N} \frac{1}{\sqrt{2\pi\sigma_i^2}} \exp\left[-\frac{1}{2}\frac{\left(d_i^{\text{pred}}(\mathbf{m}) - d_i\right)^2}{\sigma_i^2}\right], \qquad (5)$$

where $N$ denotes the number of data points and $\sigma_i$ represents the measurement error standard deviation of the $i$-th data point.

We apply a hierarchical Bayes scheme (e.g., Malinverno and Briggs, 2004) and estimate a global relative error level $\sigma_{\text{rel}} = \sigma_i / d_i$, jointly with the parameters **m**. By doing so, we also account for epistemic (model structural) and model parameterization errors, as long as these are independent and well described by the Gaussian likelihood function in eq. (5) (e.g., Schoups et al., 2010; Rosas-Carbajal et al., 2014). Relative errors are used herein, but the perhaps more common assumption of absolute errors would also be straightforward to implement and lead to similar results. The posterior pdf is sampled with the DREAM$_{(ZS)}$ algorithm (ter Braak and Vrugt, 2008; Vrugt et al., 2009; Laloy and Vrugt, 2012). We here give a brief description of the sampling scheme and refer to Laloy and Vrugt (2012) for a detailed description of the algorithm. In short, DREAM$_{(ZS)}$ is an adaptive MCMC algorithm that runs $K$ ($K > 2$) different chains in parallel and creates jumps in each chain using a fixed multiple of the difference of two states sampled from an archive of past model states. This archive is growing progressively as the sampling evolves and diminishing adaptation ensures convergence to the appropriate target distribution. We refer to ter Braak and Vrugt (2008) for an in-depth discussion concerning ergodicity and convergence properties of this algorithm. If the position of the $i$-th chain is given by $\mathbf{m}^i$, then new proposal models, $\mathbf{m}^i_{\text{prop}}$ are calculated using

$$\mathbf{m}^i_{\text{prop}} = \mathbf{m}^i + \Delta^i. \tag{6}$$

The DREAM$_{(ZS)}$ algorithm implements subspace sampling in which only a subset of all dimensions is periodically updated. The dimensions to be updated (indexed $j$) are determined for each step based on a series of crossover values (details given by Vrugt et al., 2009). The number of updated dimensions, $d'$, lies within 1 and the total number of dimensions, $d$. The respective dimensions are updated by a proposal jump

$$\Delta^i_j = \left(\mathbf{1}_{d'} + \mathbf{e}_{d'}\right)\gamma_{d'}\left[\mathbf{z}^{r1}_j - \mathbf{z}^{r2}_j\right] + \boldsymbol{\varepsilon}_{d'}, \tag{7}$$

where $\mathbf{1}_{d'}$ denotes a unit vector of length $d'$, $\mathbf{z}^{r1}$ and $\mathbf{z}^{r2}$ are sampled from the external archive of model states, $\mathbf{Z}$, and $r_1$ and $r_2$ are randomly chosen members of $\mathbf{Z}$. The variables $\mathbf{e}_{d'}$ and $\boldsymbol{\varepsilon}_{d'}$ add stochastic fluctuations to ensure ergodicity and are drawn from $U_{d'}(-b,b)$ and $N_{d'}(0,b^*)$, respectively, where $b$ and $b^*$ are small compared to the width of the target distribution. The jump rate, $\gamma_{d'} = 2.4/\sqrt{2d'}$, depends on the number of updated dimensions $d'$ (Vrugt et al., 2009) and controls the dissimilarity between subsequent states. Each fifth proposal, the jump rate is temporarily set to 1 to facilitate direct jumps between disconnected posterior modes.

The acceptance probability of the proposed model is determined using the Metropolis ratio (e.g., Mosegaard and Tarantola, 1995)

$$\alpha = \min\left\{1, \frac{p\left(\mathbf{m}^i_{\text{prop}} \middle| \mathbf{d}\right)}{p\left(\mathbf{m}^i \middle| \mathbf{d}\right)}\right\}. \tag{8}$$

If the posterior probability of the proposed model, $p\left(\mathbf{m}^i_{\text{prop}} \middle| \mathbf{d}\right)$, is larger than that of the current position of the chain, $p\left(\mathbf{m}^i \middle| \mathbf{d}\right)$, the probability to move to this new state is one. Otherwise the acceptance probability is given by the ratio of

$p\left(\mathbf{m}_{\text{prop}}^{i}|\mathbf{d}\right)$ and $p\left(\mathbf{m}^{i}|\mathbf{d}\right)$. For the sake of numerical stability, the likelihood function and the prior probability are evaluated in log-space (denoted $l(\mathbf{m}|\mathbf{d})$ and $\omega(\mathbf{m})$, respectively), and the acceptance probability to move from one state to the next becomes

$$\alpha = \min\left\{1, \exp\left(l\left(\mathbf{m}_{\text{prop}}^{i}|\mathbf{d}\right) + \omega\left(\mathbf{m}_{\text{prop}}^{i}\right) - l\left(\mathbf{m}^{i}|\mathbf{d}\right) - \omega\left(\mathbf{m}^{i}\right)\right)\right\}. \tag{9}$$

## 2.3 Summary Statistics from Training Image Realizations

Training images (TI) are 2- or 3-D images, digitally created or drawn by hand, that represent conceptual geological information of the area of interest such as characteristic structural elements and lithofacies (Strebelle, 2002; Hu and Chugunova, 2008). Such images are typically used when the subsurface is described by repeating structures, for example, meandering and intersecting channels in the case of fluvial or turbiditic reservoirs. In geostatistical modeling, the purpose of a TI is to represent the spatial dependencies between a set of points larger than 2, that is, the multiple-point statistics (MPS).

Our method relies on a TI that captures the expected spatial patterns and lithofacies information at the study site, which we refer to as geological prior information. Each lithological unit of the TI is assigned a different porosity value using information from outcrop data or borehole cuttings. Then, an ensemble of different geostatistical realizations of the TI is generated using the simulation algorithm *DeeSse*, an improved and commercialized version of the original algorithm proposed by Mariethoz et al. (2010b). *DeeSse* simulations are based on direct sampling of patterns from the TI to sequentially simulate all pixel values in the realization. Patterns are defined by a set of neighboring pixels that have already been assigned values in previous simulation steps. The pixel combination defined by the pixel values and lags of the neighboring pixels to the target pixel, that is, the pixel to be simulated, builds a pattern. The TI is scanned for such a pattern and once found, the pixel values of the TI are copied to the realization. This pixel-wise copy-paste scheme preserves the multiple-point statistics of the TI.

The size of the neighborhood (number of adjoining pixels used) determines the order of spatial statistics of the TI that is honored in each realization. Indeed, if only one neighbor is used, this procedure is equivalent to regular variogram sampling. Simulation quality and computational cost of the *DeeSee* algorithm depend on a set of controlling algorithmic parameters, which we chose following the recommendations made by Meerschman et al. (2013). Of course, the ensemble of TI realizations should be large enough to ensure stable and thus reliable summary statistics.

To compress the model space, each TI realization is transformed into the discrete cosine domain and the dominant transform coefficients are determined (see Section 2.1). The DCT of the TI realizations are then truncated by setting all coefficients to zero except the dominant ones (note that these dominant coefficients are representative of all TI realizations). The inverse DCT of these truncated spectra yields an ensemble of porosity realizations. These porosity realizations, rather than their original counterparts, are used to extract summary metrics. In this way, we ensure that the summary statistics describe features that can be represented by the sparse model parameterizations. The various steps of this procedure are shown in Fig. 1.

To illustrate our methodology, we use three different summary metrics of model morphology: (a) overall model roughness, (b) parameter variability, and (c) porosity percentiles. These three metrics summarize important information about the expected spatial distribution of porosity. Furthermore, they are easy to calculate and invoke only minimal computational cost.

The roughness of the model, $S_r$, is calculated from the model representation on a Cartesian grid, **A**, using first-order differences in the *x*- and *z*-direction

$$S_r = \sum_{x=2}^{N_x} \sum_{z=2}^{N_z} \left( \left| A(x,z) - A(x-1,z) \right| + \left| A(x,z) - A(x,z-1) \right| \right). \tag{10}$$

Such a roughness term is, in a least-squares sense, part of the objective function in smoothness-constrained deterministic inversions, where the goal is to find least-structured models that explain the data (e.g., Constable et al., 1987). In these approaches any measure of structure larger than zero is penalized, thereby resulting in models that are often overly smooth. Here, we use the TI realizations to extract a distribution describing the expected roughness. A global measure of roughness similar to $S_r$ is often imposed as a constraint in image processing, where it is commonly referred to as total variation (TV, e.g., Rudin et al., 1992).

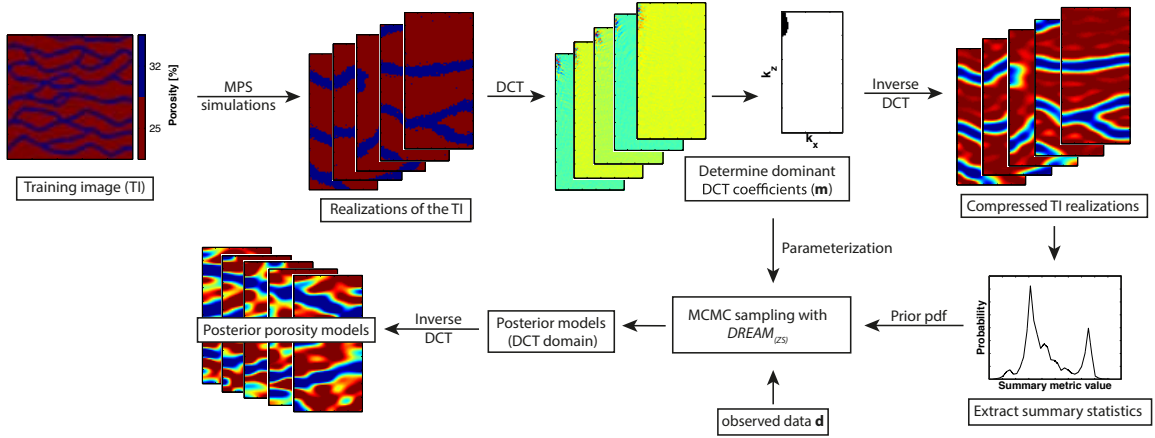

**Figure 1:** Schematic workflow of the presented inversion strategy. Realizations of a training image (TI) generated by multiple-point statistics (MPS) simulations are subject to discrete cosine transformation (DCT). Dominant DCT coefficients are derived and the corresponding basis functions are used to parameterize the porosity fields. From compressed TI realizations, summary statistics distributions are extracted and translated into a prior probability density function. The model space is sampled with the $DREAM_{(ZS)}$ algorithm to obtain posterior model realizations.

Another summary metric used herein is the total sum of the absolute values of DCT coefficients used to parameterize the model

$$S_v = \sum_{k_x=1}^{N_x} \sum_{k_z=1}^{N_z} \left| B_{\text{trunc}}(k_x, k_z) \right|, \tag{11}$$

where $\mathbf{B}_{\text{trunc}}$ is the sparse model representation in the DCT domain. This metric measures the expected variability, or energy, within the model image as $S_v$ is large for models with many interfaces and strongly varying porosities at different spatial scales. The TI realizations thus allow us to extract explicit information about the expected variability.

To make sure that our posterior models exhibit the "right" porosity distribution, we consider as third, and last summary metric namely the 10, 50 and 90% percentiles of the porosities from our TI realizations. We summarize these values in the vector $\mathbf{S}_p = (S_{10}, S_{50}, S_{90})$.

## 2.4 Prior Distribution Based on Summary Statistics

Our model vector contains the *n* dominant DCT coefficients that define the porosity field, a set of petrophysical parameters that relate the porosity to GPR wavespeed values, and the relative error level $\sigma_{\text{rel}}$. All model parameters are initially sampled from uniform distributions. The ranges of the retained DCT coefficients are set equal to their maximum and minimum values found in the ensemble of TI realizations. Further, the upper and lower bounds of the petrophysical parameters are based on literature values (Keller, 1988). We assume that the relative error level follows a Jeffreys prior (i.e., its logarithm is uniformly distributed) with a range of half and two times its true value (Tarantola, 2005).

The premise of this paper is that summary metrics from TI realizations can help to constrain model morphology. The idea is to translate the distribution of these summary metrics observed in the TI realizations into a prior distribution. This can thus be seen as an indirect prior on the model parameters, as the summary metrics of a model depend on the porosity values which in turn depend on the sampled DCT coefficients and petrophysical parameters.

To handle prior distributions of arbitrary shape, we create histograms of the summary metrics of the TI realizations. The frequencies on the *y*-axis are scaled so that the histogram integrates to unity, and represents a proper probability distribution. The prior probability of a proposal model, $p(\mathbf{m}_{\text{prop}})$, is in the case of independent summary statistics given by

$$p(\mathbf{m}_{\text{prop}}) = \prod_{j=1}^{d_S} p_{j,k}, \tag{12}$$

where $d_S$ signifies the dimensionality of **S** and $p_{j,k}$ is the probability of the *k*-th bin for the *j*-th summary metric. The assumption of independence among the summary metrics is acceptable for the studies considered herein (e.g., correlation coefficients smaller than 0.1 for the first case study) but might not be reasonable for other problems and a multivariate prior distribution should be used.

This distribution constitutes a bounded indirect prior, as the bins are restricted to a certain range and the probability assigned to each bin refers to the summary metric, not to the sampled individual model parameters. This, however, does not guarantee that the summary metrics of the proposed models will fall into the appropriate range covered by this prior. For example, consider Fig. 2 that presents a histogram of the sampled model roughness values for one of the case studies discussed later. The gray bars depict the roughness distribution in an ensemble of 1,000 TI realizations. Each bin is assigned a probability $p_k$ (black solid line). The roughness values of the models from the initial archive **Z** used in DREAM$_{(ZS)}$ are shown in blue. None of these models are deemed acceptable, as their roughness values fall outside the prescribed prior distribution for this summary metric. This happens as DREAM$_{(ZS)}$ in the initiation stage draws independent realizations of all model parameters (i.e., DCT coefficients and petrophysical parameters) without any consideration of the summary statistics.

To ensure that all models honor the observed summary metrics we draw inspiration from Sadegh and Vrugt (2014) and use an indicator function on the prior distribution. This function $I(\mathbf{m})$ returns 1 if the summary metrics within a proposed model realization is within the range of the observed summary metrics derived from the TI realizations. Outside this range, $I(\mathbf{m})$ decreases linearly to 0 at the extreme ends. The acceptance rule to move from one model state to the next is then modified such that $\mathbf{m}^i_{\text{prop}}$ is always accepted if $I(\mathbf{m}^i_{\text{prop}}) > I(\mathbf{m}^i)$, otherwise the proposal is rejected unless $I(\mathbf{m}^i_{\text{prop}}) = 1$. In this latter case, the acceptance ratio is calculated following eq. (9). The range over which the indicator function is defined is determined by the minimum and maximum values of the summary metrics observed in the combined set of TI realizations and initial archive of models, **Z**. If several summary metrics are considered, the indicator values are multiplied, meaning that all summary metric values must lie in the prior range for the proposal to be evaluated as a possible posterior sample.

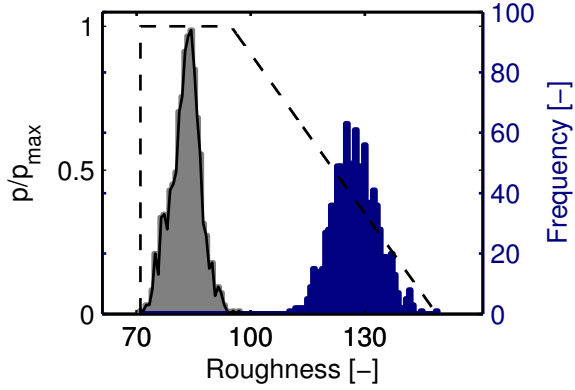

**Figure 2:** Formulation of the summary metrics prior and the indicator function. The distribution of roughness values observed in a set of 1,000 TI realizations is shown in gray and the estimated prior following this distribution is shown by the black solid line. The blue bars depict the distribution of roughness values in an initial set of random models. To ensure that the posterior models have roughness values within the summary metric prior range, an indicator function (dashed black line) is applied. The indicator function is 1 where the summary metrics prior is non-zero and it decreases to 0 towards the possible extreme values that might be sampled. Only realizations for which the indicator function is 1 are considered as possible realizations of the posterior.

As pointed out by Sadegh and Vrugt (2014), this modified acceptance rule introduces two different search stages of DREAM$_{(ZS)}$. Initially, the algorithm acts as an optimizer and only accepts proposals if the summary metrics of the proposals are in better agreement with their observed values (smaller Euclidean distance). The resulting transitions of the Markov chain are irreversible, and hence the sampled states in stage 1 cannot be used to infer statistical properties of the model parameters. Once the simulated summary metrics are within their desired prior range ($I(\mathbf{m}_{\text{prop}}^i) = 1$), stage 2 is initiated and equation (9) is used to determine whether to accept each proposal or not. This then constitutes a regular MCMC step, and should, after sufficient burn-in, provide a sample of the posterior target distribution. In stage 2, it regularly happens that the proposal jump (eq. 7) leads to models with summary statistics that are outside of the desired prior range. When this happens, new proposals are generated with eq. (7) until $I(\mathbf{m}_{\text{prop}}^i) = 1$. The acceptance probability of eq. (9) then determines whether this proposal is accepted, otherwise the chain remains at its current state. Empirical results presented herein illustrate that this two-stage acceptance rule works well in practice.

The summary statistics have been constructed from a population of TI realizations and can be represented with our model parameterization. This guarantees the presence of a region within the model space which satisfies the observed summary metrics. However, this does not discount the possibility of multiple disconnected behavioral regions that each honor the observed summary metrics. Such disconnected posterior distributions are very difficult to sample adequately. Fortunately, by using $K = 5$ different chains in DREAM$_{(ZS)}$ the model space is exhaustively explored, and by setting the jump rate periodically to unity in eq. (7), the chains can jump directly from one mode to the next. This enables inference of complex and multimodal posterior distributions (Vrugt et al., 2009, Laloy and Vrugt, 2012).

## 2.5 Forward Modeling

The proposed inversion method is applied to first-arrival travel time data from synthetic crosshole GPR experiments. A radar wave is transmitted in one borehole and recorded in an adjacent one. By measuring the first-arrival time of the signal for various transmitter-receiver configurations, we can derive information about the spatial distribution of the radar slowness $u$ (i.e., the reciprocal of the wavespeed) between the boreholes, which for a water-saturated porous medium can be related to the 2-D porosity distribution $\varphi(x,z)$ through the following petrophysical relation (Pride, 1994; Davis and Annan, 1989)

$$u(x,z) = \frac{1}{c}\sqrt{\varphi(x,z)^m (\varepsilon_w - \varepsilon_s) + \varepsilon_s}, \tag{13}$$

where $c$ [m/s] signifies the speed of light in a vacuum, $m$ [-] denotes the cementation factor, and $\varepsilon_w$ [-] and $\varepsilon_s$ [-] are the relative electrical permittivities of water and grains, respectively. In our analysis, we assume that $c = 3 \times 10^8$ m/s and $\varepsilon_w = 81$, and consider $m \in [1.3, 1.8]$ and $\varepsilon_s \in [2,6]$ to be free parameters in the inversion. The ranges of $m$ and $\varepsilon_s$ are taken quite wide and are based on literature values (Keller, 1988). We refer to Peterson (2001) for a detailed discussion about suitable field procedures when acquiring crosshole GPR data and an assessment of the sources of errors that affect the picked travel times.

The forward problem thus involves the calculation of the GPR travel times between the two boreholes, which is a function of the unknown porosity field and petrophysical relationship. We obtain the spatial distribution of the first-arrival travel times for each source position by solving the eikonal equation using a finite-difference scheme (Podvin and Lecomte, 1991).

## 3 Results

To benchmark our inversion method, we use two synthetic case studies. The first study involves a subsurface of channel structures within a homogeneous matrix. This results in a binary TI. The second study involves a more complex geological setting featuring five lithological units forming typical sedimentary structures of a fluvial deposit. Both cases are described in detail by Lochbühler et al. (2014b). The TIs for both studies and a few of their corresponding TI realizations are shown in Fig. 3. A total of 1,000 realizations were generated from each of the TIs and compressed to $n = 100$ and $n = 150$ DCT coefficients, respectively (see section 2.1). For both case studies we used standard settings of the algorithmic variables of DREAM$_{(ZS)}$ (Laloy and Vrugt, 2012). Convergence was assessed using the $\hat{R}$-statistic of Gelman and Rubin (1992). This statistic compares for each parameter of interest the between- and within-variance of the chains. Because of the asymptotic independence, the between-member variance and $\hat{R}$ can be estimated consistently from a single DREAM$_{(ZS)}$ trial. In practice, values of $\hat{R}$ smaller than 1.2 indicate convergence to a limiting distribution. In all our calculations reported herein we use the last 50% of the samples in each chain to calculate the $\hat{R}$ diagnostic. Simulation results show that about 40,000 and 70,000 iterations were required to reach convergence for case study 1 and 2, respectively.

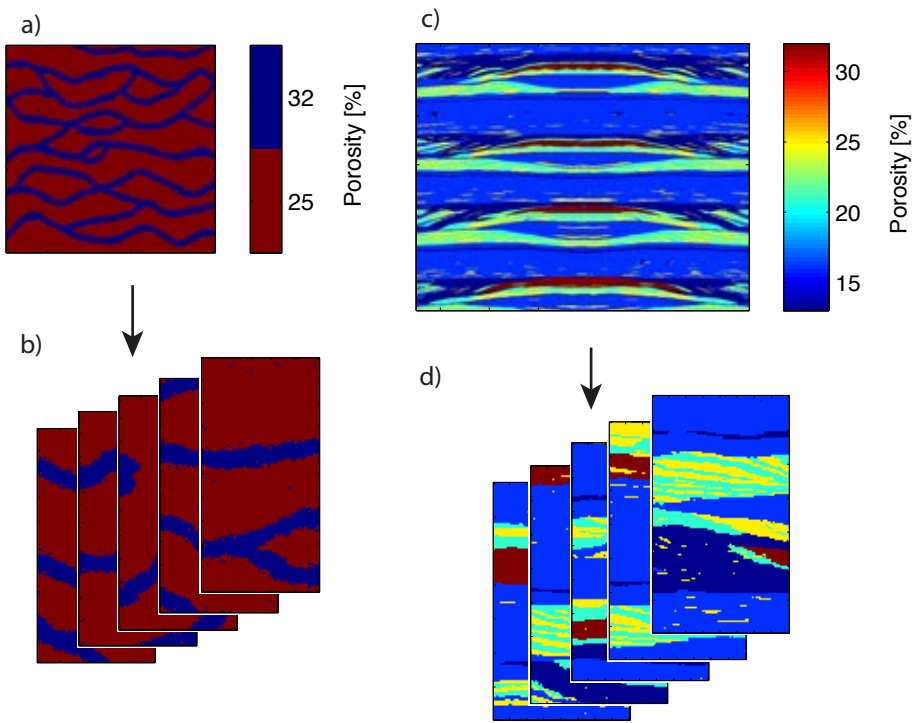

**Figure 3:** Training images for the channels case (a), the fluvial deposits case (c) and multiple-point statistics realizations thereof (b and d). Note that the full ensemble of TI realizations contains 1,000 realizations for both cases.

## 3.1 Channels Case

Our first study, hereafter referred to as the channels case, considers channels of porosity $\varphi = 25\%$ that pervade a matrix with $\varphi = 32\%$. The relative electrical permittivity of the grains is $\varepsilon_s = 3$ and the cementation factor, $m$, is 1.5. The forward problem is solved on a grid with square blocks of 0.1 × 0.1 m. GPR transmitter and receiver antennas are placed at a distance of 0.3 m along the left and right boundary of the domain (c.f. white dots in Fig. 4a), resulting in a total of 879 travel times. The simulated travel times are corrupted with a heteroscedastic Gaussian measurement error with standard deviation set to 2% of the 'observed' values. A relative error is used here to partly account for the typically larger observational and modeling errors associated with high-angle ray paths that are often associated with longer travel times, but similar results are expected if using absolute errors. The reference porosity field and its compressed image are shown in Figs 4(a) and (b), respectively. As we use a sparse model parameterization, the compressed image is, in a visual sense, the 'best' model we can theoretically recover.

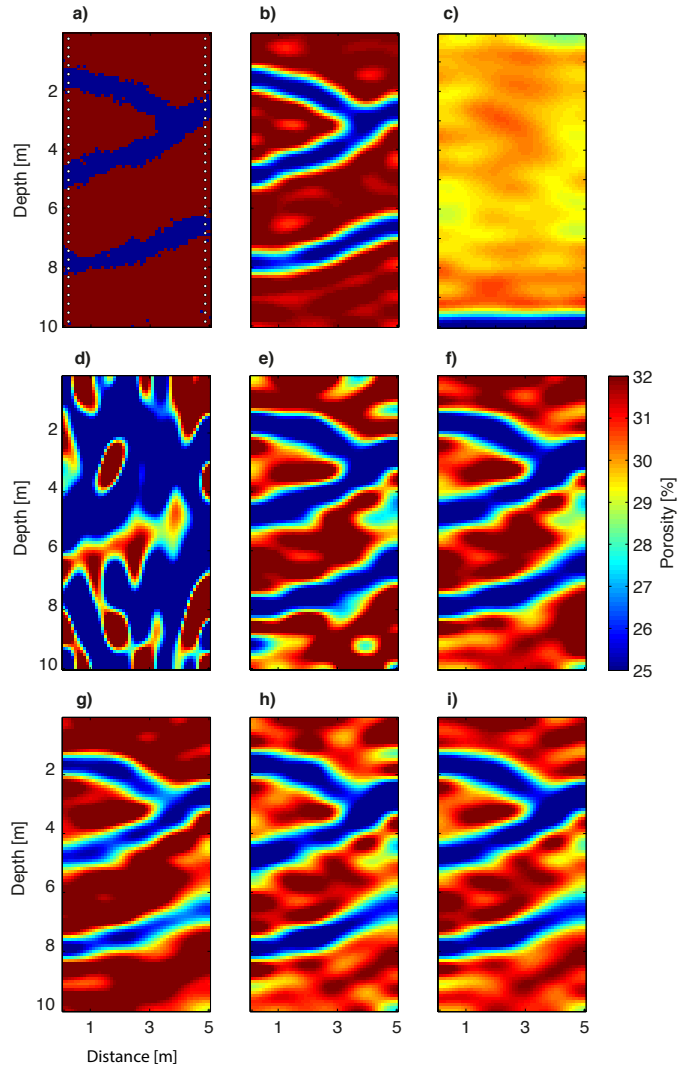

**Figure 4:** Channels case: a) reference porosity field with GPR transmitter (left side) and receiver (right side) antenna locations indicated by white dots, b) compressed image of the reference porosity field obtained by retaining the *n* = 100 highest amplitude DCT coefficients of the reference field. c) Ensemble mean of the summary metrics prior (variability and porosity percentiles) using the TI-based parameterization obtained by running an MCMC simulation with the likelihood set to unity. d) and e) Mean of the inferred porosity posterior without consideration of a summary metrics prior without (d) and with (e) TI-based parameterization. f)-i) Posterior mean of models obtained with TI-based parameterization and considering a summary metrics prior using f) roughness metric, g) variability metric, h) porosity percentile metrics and i) both variability and percentile metrics.

We first draw random simulations of our summary-statistics based prior using the TI-based parameterization by running the MCMC algorithm with a likelihood of unity. Eight sample realizations (Fig. 5a) suggest that the summary-statistics based prior favors correct proportions between high- and low-porosity regions, but that the low-porosity regions (channel material) are generally disconnected. The ensemble mean (Fig. 4c) has no resemblance with the reference model (Figs. 4a-b), but the patterns are not truly random. This is a consequence of our model parameterization strategy. It is clear that any channel-like structures that appear in the posterior model realizations are the consequence of conditioning to the geophysical data. To illustrate the effect of the chosen model parameterization, the data were first inverted without a summary metrics-based prior distribution or a TI-based model parameterization. Instead, all but the 100 low-frequency coefficients arranged in a 10 × 10 rectangle were set to zero (Linde and Vrugt, 2013; Lochbühler et al., 2014a). Even though the data are almost fitted up to their measurement error, the inverse models bear very little resemblance with the reference porosity field (Figs. 4d and 5b). With a TI-based model parameterization (section 2.1), the inversion results are markedly improved (Figs 4e and 5c) in that the main structural features of the reference porosity field are captured. However, inversion artifacts are abundant and $m$ and $\varepsilon_s$ are poorly estimated (Figs 5c, 6b). The use of a summary statistics-based prior distribution further improves the posterior models. Many of the inversion artifacts persist and the petrophysical parameters are much closer to their true values (Figs 6c-f), particularly if the prior distribution is based on the observed sum of DCT coefficients (Fig. 5e).

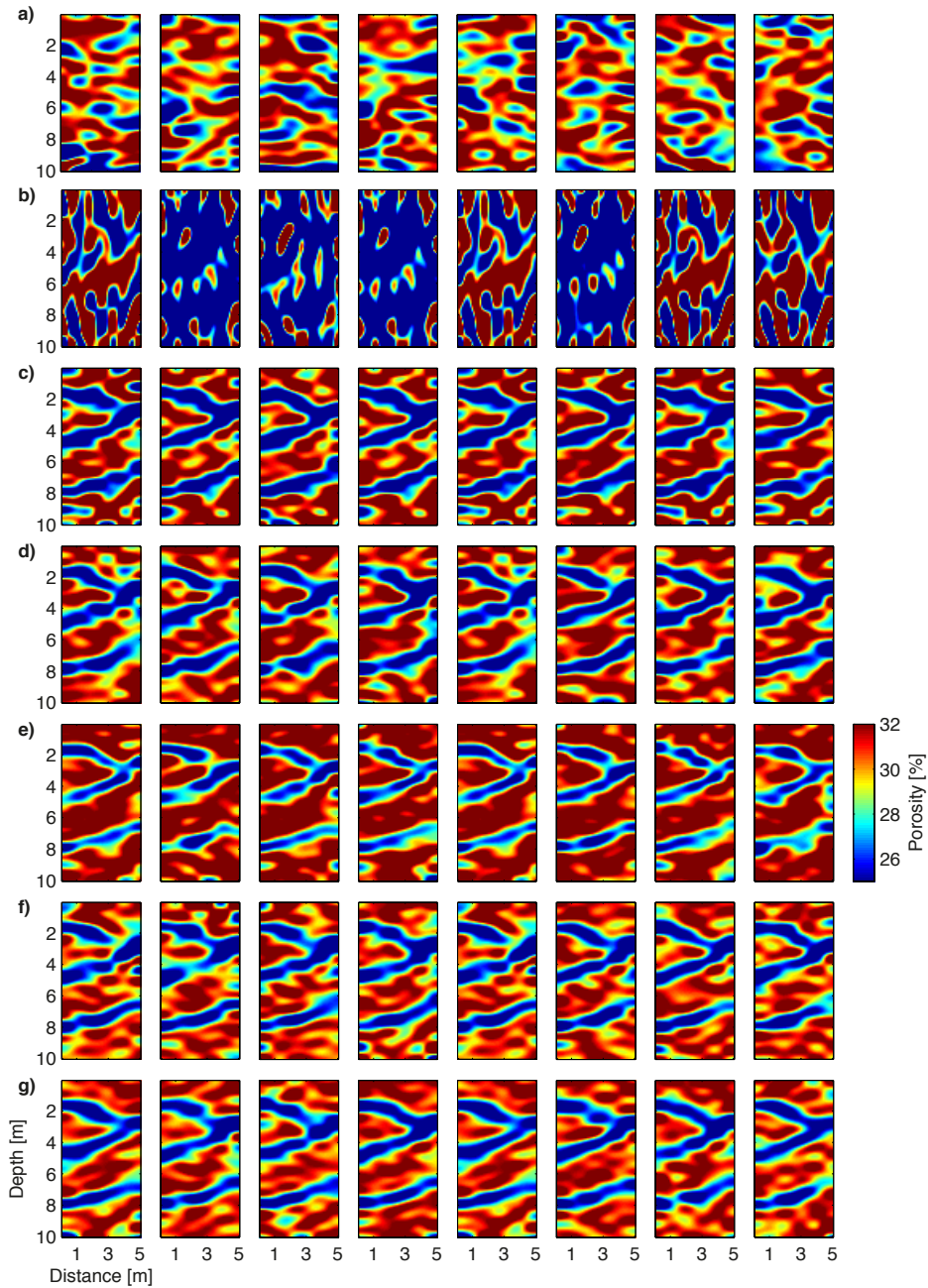

**Figure 5:** a) Random samples from the ensemble of summary metrics prior (variability and porosity metrics) models using the TI-based parameterization for the channels case. Corresponding random samples from the ensemble of posterior models for the channels case: b) no TI-based parameterization, no summary metric prior, c) TI-based parameterization, no summary metric prior, d)-g) TI-based parameterization, summary metric prior using d) roughness metric, e) variability metric, f) porosity percentile metrics, g ) both variability and percentile metrics.

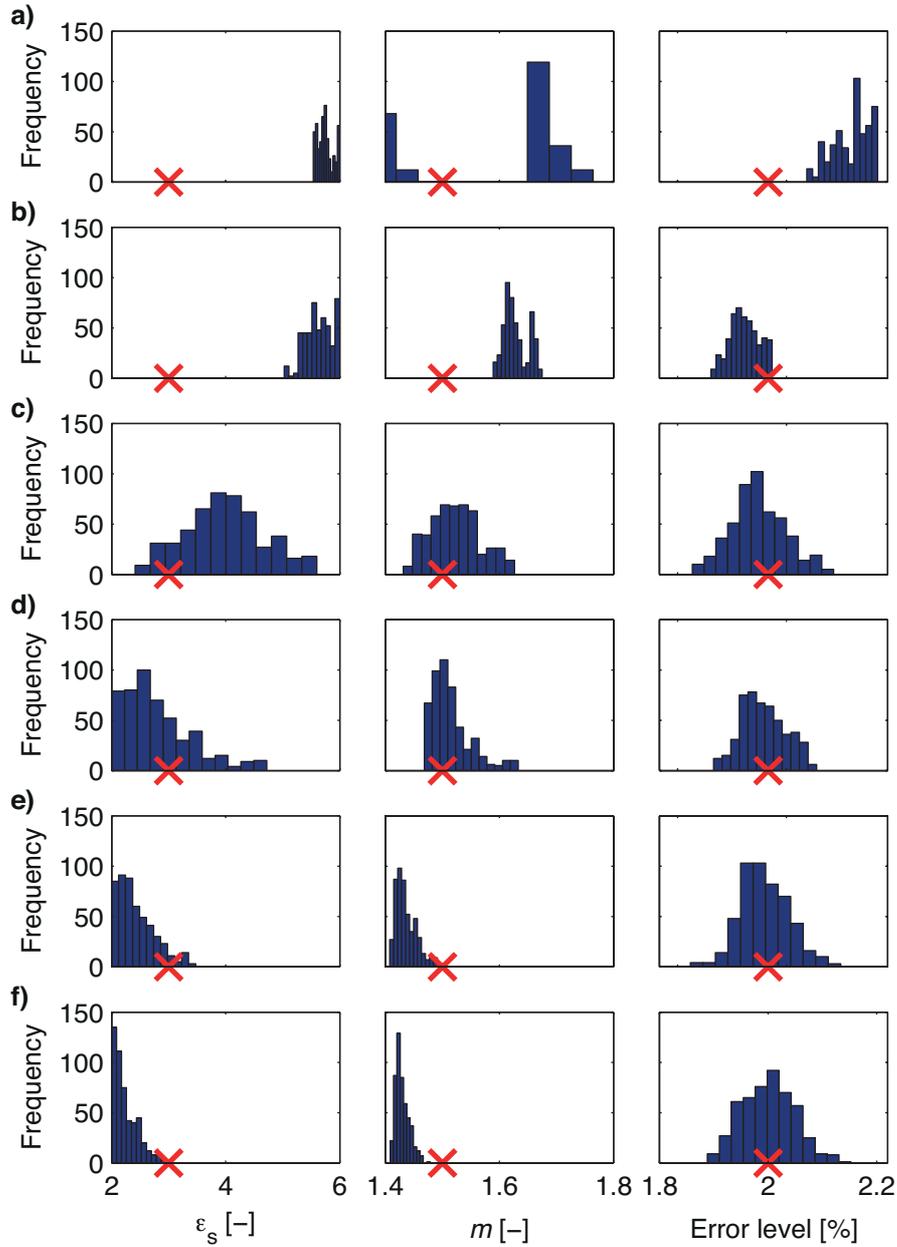

**Figure 6:** Posterior distributions of the petrophysical parameters $\varepsilon_s$ and $m$, and of the relative data error level for the channels case: a) no TI-based parameterization, no summary metric prior, b) TI-based parameterization, no summary metric prior, c)-f) TI-based parameterization, summary metric prior using c) roughness metric, d) variability metric, e) percentile metrics, f) both variability and percentile metrics. Red crosses depict the reference values.

To quantify the quality of the posterior models, we calculated the root mean square error (RMSE) between the reference porosity field $\mathbf{A}_{\text{ref}}$ (c.f., Fig. 4a) and the posterior porosity models, $\mathbf{A}$,

$$\text{RMSE} = \left\| \frac{1}{\sqrt{N_x N_z}} (\mathbf{A}_{\text{ref}} - \mathbf{A}) \right\|_2. \tag{14}$$

As shown in Table 1, the posterior models provide better porosity estimates if a TI-based parameterization is applied and the results are further improved if summary statistics are used. The closest resemblance to the reference model is achieved by joint use of the variability and the percentile metrics (mean RMSE=2.27%).

**Table 1:** Mean and standard deviation of the average root mean square error between posterior porosity models and the reference model for different degrees of prior information.

| Considered prior information | mean RMSE [%] | std of RMSE [%] |
|---|---|---|
| Channels case | | |
| no TI-based parameterization, no summary metric | 9.37 | 0.84 |
| TI-based parameterization, no summary metric | 3.30 | 0.14 |
| TI-based parameterization, roughness metric | 2.57 | 0.10 |
| TI-based parameterization, variability metric | 2.36 | 0.10 |
| TI-based parameterization, percentile metric | 2.45 | 0.11 |
| TI-based parameterization, variability and percentile metrics | 2.27 | 0.10 |
| Fluvial deposits case | | |
| no TI-based parameterization, no summary metric | 12.30 | 1.20 |
| TI-based parameterization, no summary metric | 5.44 | 0.63 |
| TI-based parameterization, roughness and percentile metrics | 4.20 | 0.31 |

The aberrant values of the petrophysical parameters demonstrate that it is not particularly easy to find "physically realistic" models in high-dimensional inversion. There are many models for which the data are adequately explained even if the petrophysical parameters are poorly estimated. This finding is perhaps not surprising, as the DCT parameterization is flexible enough to correct for this. The 2% measurement error is contained in the posterior distribution for all the different error level estimates (Fig. 6, rightmost column), except for the inversion with no TI-based parameterization (Fig. 6a).

To further demonstrate the benefits of the TI-based model parameterization and the summary metrics prior, we calculated the log posterior probability log($p(\mathbf{m}|\mathbf{d})$), with $m$ and $\varepsilon_s$ varying between their prior bounds and the DCT coefficients kept fixed at their maximum a-posteriori density (MAP) values (Fig. 7). The TI-based parameterization does not have a strong effect on the posterior density of the petrophysical parameters (compare Figs. 7a-b), but the results are quite different when the summary statistics-based prior is used (Figs. 7c-f). The MAP values (black crosses) are now in closer agreement with the $m$ and $\varepsilon_s$ values used to generate the synthetic GPR data (compare black and red crosses in Fig. 7). The results not only favor the use of summary metrics, but also demonstrate the ability of the DREAM$_{(ZS)}$ algorithm, with modified acceptance rule, to locate and sample the underlying target distribution. Indeed, the MAP values of the petrophysical parameters (blue asterisks) coincide with the location of the posterior maximum (black crosses).

To provide more insights into the sampled posterior models, consider Fig. 8 that plots prior and posterior distributions of the five different summary metrics used herein. The summary metrics of the posterior models lie exactly within their respective prior distribution (as expected), yet, they congregate in regions with relatively low prior probability (see Figs 8a and d). Nevertheless, the indicator function equals 1 (not shown), and the models are deemed behavioral. Note however, that none of the posterior models envelop the summary metrics of the reference model, except for the 90% percentile of the simulated porosity values (Fig. 8e). This highlights the dominance of the likelihood term (many data) over the prior term (few summary statistics) in calculation of the posterior density.

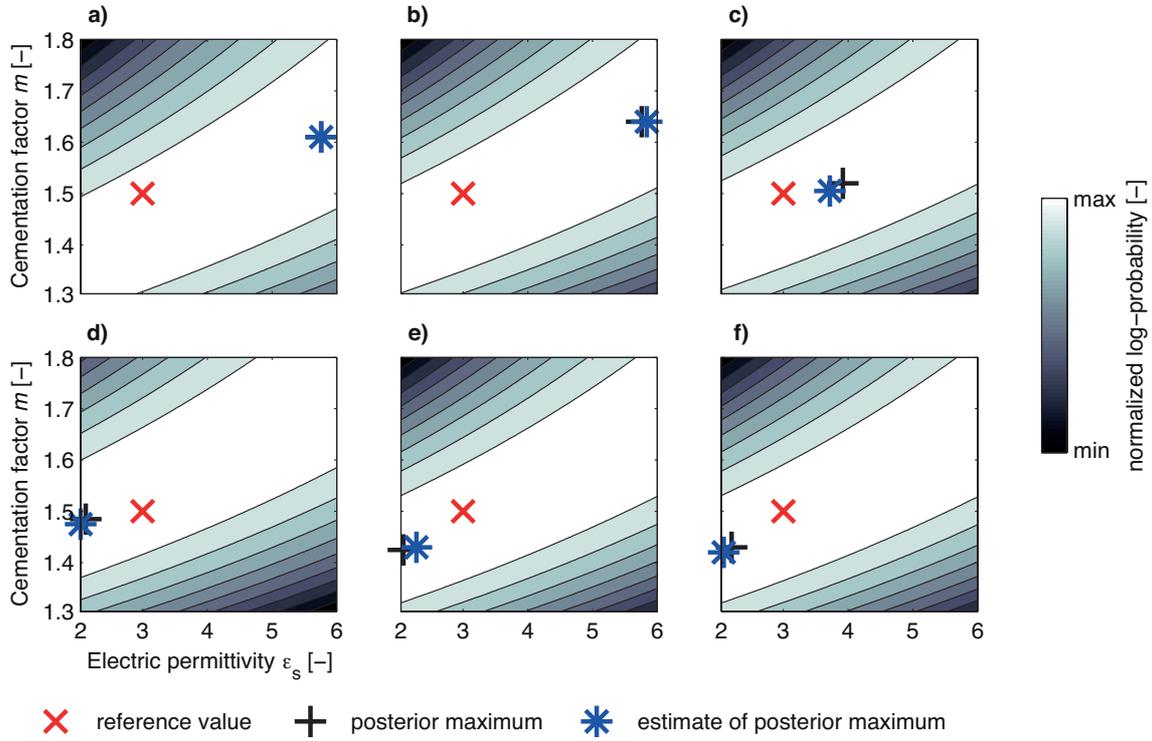

**Figure 7:** Maps of the posterior log-probability for all possible combinations of the petrophysical parameters $m$ and $\varepsilon_s$ for the channels case. The DCT coefficients are fixed at their maximum a-posteriori density (MAP) values. Indicated are the parameter values of the reference (red crosses), the values for which the log-probability is maximized (black crosses) and the estimate found by DREAM$_{(ZS)}$ (blue asterisks). Enumerations correspond to those of Fig. 6.

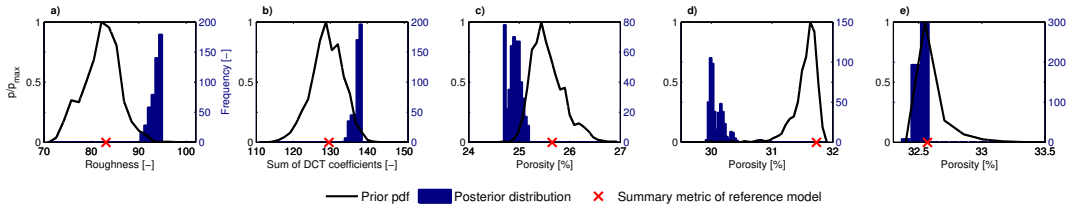

**Figure 8:** Summary metric prior distributions (black lines, see Fig. 2 for explanation) and the posterior distributions of the summary metrics (blue bars) for the channels case: a) roughness metric, b) variability metric, c)-e) 10, 50, 90% percentile metrics. Red crosses display the summary metrics of the reference porosity field.

## 3.2 Fluvial Deposits Case

The second study, hereafter referred to as fluvial deposits case, is significantly more complex (Figs 9a-b). Five lithofacies with porosities ranging between 13 and 32% create a sedimentary environment of fluvial deposits (Bayer et al., 2011; Comunian et al., 2011; Lochbühler et al., 2014b). We discretize the forward problems using blocks of 0.05 × 0.05 m and 0.25 m sensor spacing for the GPR experiments. This results in a total of 703 simulated travel times. The values of the petrophysical parameters $\varepsilon_s$ and $m$ and the relative magnitude of the data measurement error are equivalent to those used in the channels case.

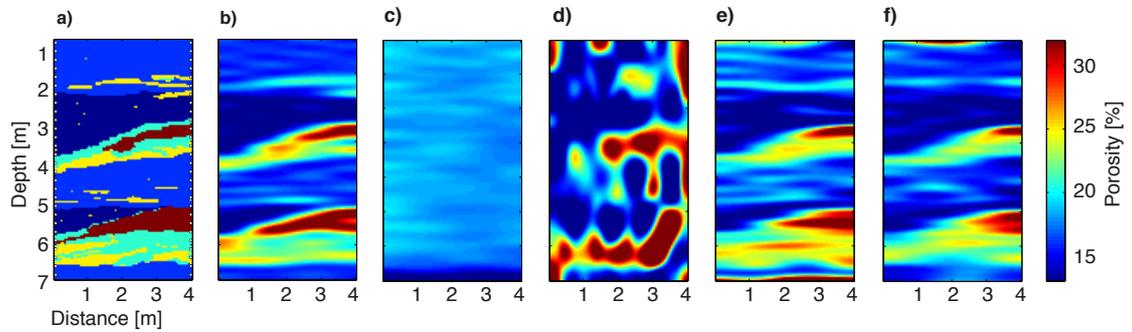

**Figure 9:** Fluvial deposits case: a) reference porosity field with GPR antenna locations indicated by white dots, b) compressed image of the reference porosity field obtained by retaining the $n = 150$ highest amplitude DCT coefficients of the reference field. c) Ensemble mean of the summary statistics prior (roughness and porosity percentiles) using the TI-based parameterization obtained by running an MCMC simulation with the likelihood always equal to 1. d) and e) Mean of the inferred porosity posterior without consideration of a summary metric prior without (d) and with (e) TI-based parameterization. f) Posterior mean of models with TI-based parameterization considering a summary metric prior based on roughness and porosity percentiles.

For this case we obtained the best results by combining the roughness and percentile metrics and we therefore restrict our analysis to those results. We first draw random simulations of our summary-statistics based prior using the TI-based parameterization by running the MCMC algorithm with a likelihood of unity. Eight sample realizations (Fig. 10a) indicate that the summary-statistics based prior favors horizontal structures and a certain degree of variability. The ensemble mean (Fig. 9c) shows little structure and provides no information about the actual location of high and low porosity regions. This information can only be obtained by considering the actual data. Inversion based on a model parameterization with a set of low-frequency DCT coefficients (here, a 15 × 10 rectangle is used) without a summary statistics-based prior does not produce adequate results (Figs 9d, 10b, 11a). The use of a TI-based parameterization leads to a much better recovery of the main structures of the reference field, and the added use of summary metrics further improves the inversion results by suppressing inversion artifacts and improving the petrophysical parameter estimates (Figs 9e,f, 10c,d, 11b,c, Table 1). The relative error level is well retrieved in all cases that consider information from the TI realizations (Figs 11b-c) and the geophysical data. The benefit of using information from TI realizations is also illustrated in Fig. 12. The MAP values are in much better agreement with their reference (true) values, particularly if summary metrics are used (Figs 12b-c).

To provide further insights into the inversion results, consider Fig. 13 that presents plots of the prior (black line) and posterior (blue histogram) distribution of each summary metric. The reference values of the summary metrics are separately indicated at the bottom of each plot with a red cross. The posterior estimates of the summary metrics are in close agreement with the corresponding values of the reference model. The percentiles of the porosity distribution are particularly well described and their marginal posterior distributions encompass the reference values.

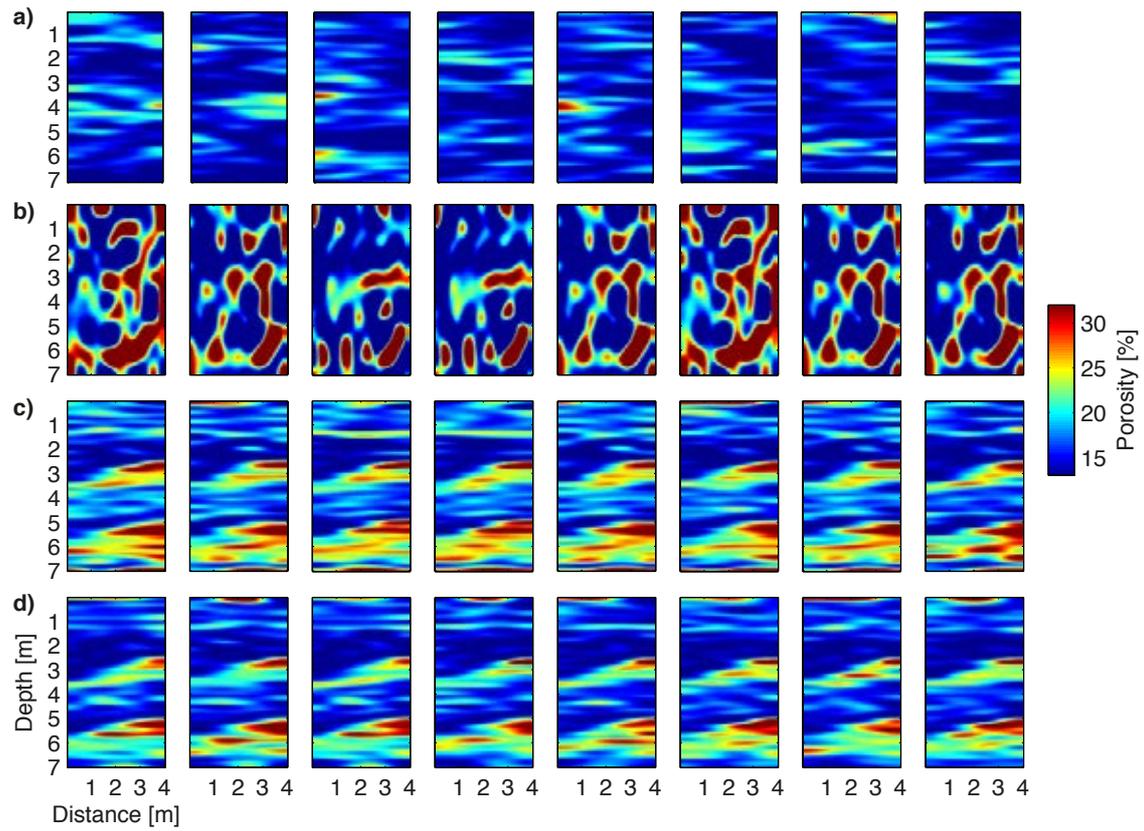

**Figure 10:** a) Random samples from the ensemble of summary statistics prior (roughness and porosity percentiles) using the TI-based parameterization for the fluvial deposits case. Corresponding random samples from the ensemble of posterior models: b) no TI-based parameterization, no summary metric prior, c) TI-based parameterization, no summary metric prior, d) TI-based parameterization, summary metric prior using roughness and porosity percentile metrics.

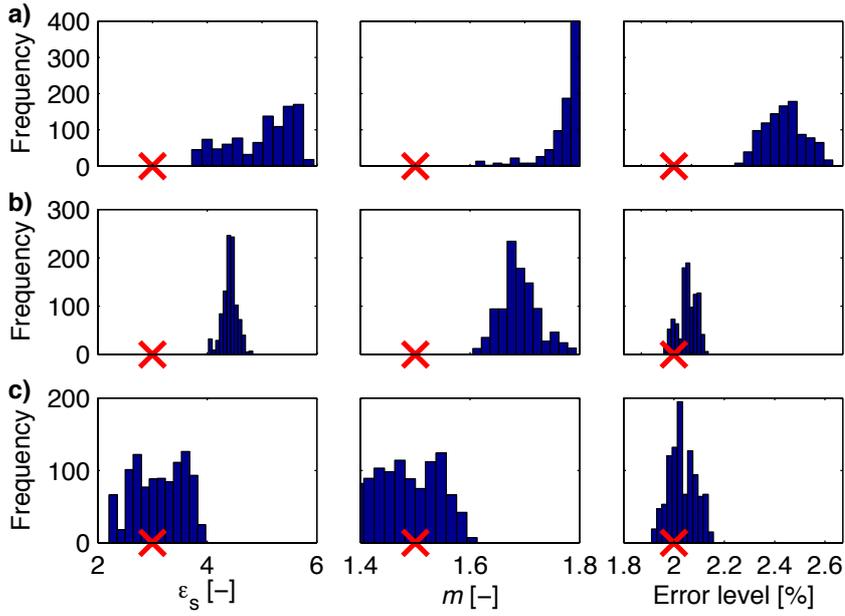

**Figure 11:** Posterior distributions of the petrophysical parameters $\varepsilon_s$ and $m$, and of the relative data error level for the fluvial deposits case: a) no TI-based parameterization, no summary metric prior, b) TI-based parameterization, no summary metric prior, c) TI-based parameterization, summary metric prior using roughness and percentile metrics. Red crosses depict the reference values.

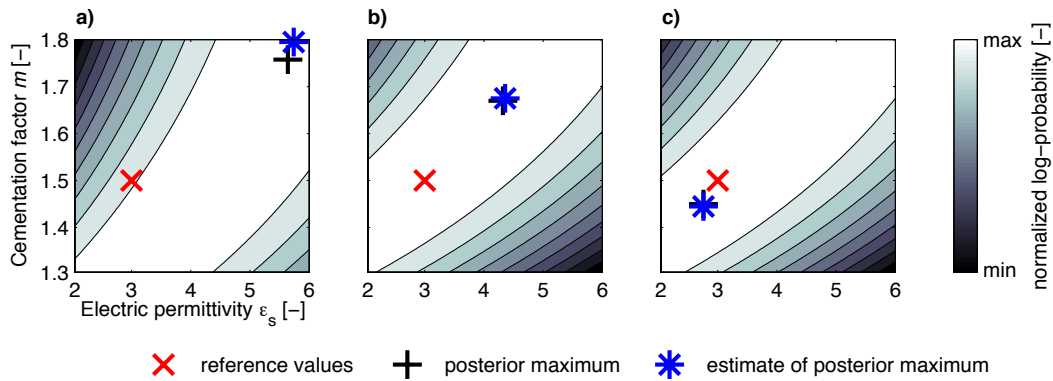

**Figure 12:** Maps of the posterior log-probability for all possible combinations of the petrophysical parameters $m$ and $\varepsilon_s$ for the fluvial deposits case. The DCT coefficients are fixed at their maximum a-posteriori density (MAP) values. Indicated are the parameter values of the reference (red crosses), the values for which the log-probability is maximized (black crosses) and the estimate found by DREAM$_{(ZS)}$ (blue asterisks). Enumerations correspond to those of Fig. 11.

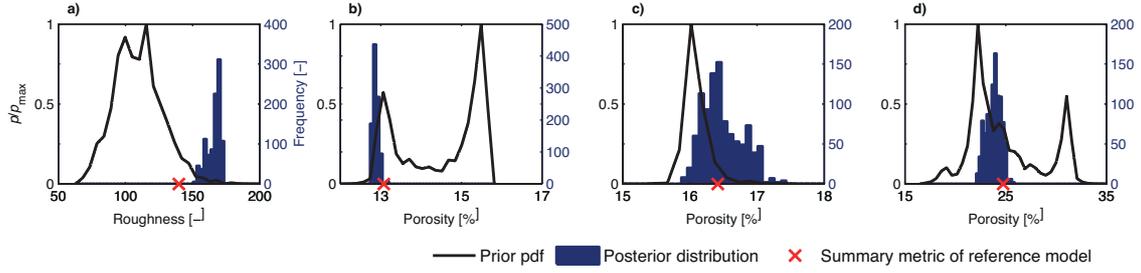

**Figure 13:** Prior probability density functions derived from the distribution of summary metrics in the ensemble of TI realizations (black lines, see Figure 2 for explanation) and the posterior distributions of the summary metrics (blue bars) for the fluvial deposits case: a) roughness metric, b)-d) 10, 50, 90% percentile metrics. Red crosses depict the summary metrics of the reference porosity field.

## 4 Discussion

We demonstrate that a TI can be used to define case-specific model parameterizations and to impose prior constraints on model morphology. This prior distribution is derived from an ensemble of TI realizations, and summarizes the frequency distribution of different summary statistics describing model morphology and lithology. The use of such prior restricts the feasible solution space, and significantly enhances the reliability of the posterior models. Applications are shown for crosshole radar tomography, but this general-purpose method can be applied to any type of data, as long as the observations are informative and sufficiently sensitive to the model parameters. In the two case studies considered herein, the summary metrics are extracted from 2-D MPS TIs, but the method is readily applicable to 3-D TIs. Furthermore, the TIs do not have to incorporate higher-order statistics. If the spatial distributions of the subsurface parameters are well-described by a multivariate Gaussian or other two-point statistical dependencies, and if these dependencies can be expressed as summary metrics, then it is straightforward to incorporate them in our methodology. The use of an indicator function enforces the posterior summary metrics to lie within the range derived from the TI realizations. This constitutes an optimization step before the posterior is reached (Sadegh and Vrugt, 2014), and hence violates detailed balance. As soon as the indicator function is unity, and the simulated summary metrics lie within their respective prior distribution, the chain becomes Markovian as the indicator function is no longer informative, and the posterior distribution solely depends on the likelihood and prior probability.

The presented inversion framework can accommodate a wide range of summary statistics. The inversions presented herein are constrained by summary metrics describing the expected roughness, parameter variability, and percentile porosity distribution. The summary metrics used in each of the case studies significantly improved the inversion results, not only in terms of visual agreement with the reference model but also with respect to the posterior parameter estimates. Of course, the usefulness of each summary metric is case study dependent. The decision which summary metrics to use depends on the outcome of the analysis of the TI realizations. If, for instance, the prior distribution of a certain summary metric is flat then its use in the inversion would seem rather unproductive. On the contrary, metrics that are well-defined, are expected to help constrain the posterior models. For illustrative purposes, we have used rather simple and common metrics. If deemed appropriate, many other summary metrics can be defined and used, for example those that measure internal coherency (McClymont et al., 2008), preferred lithological orientation (e.g., Chugunova and Hu, 2008), connectivity (e.g., Renard et al., 2011), sparseness (e.g., Jafarpour et al., 2009), etc. More advanced summary metrics could be related to MPS specific information. Recently, Lange et al. (2012) constrained simulated annealing optimization with the pattern frequency from an MPS TI. Honarkhah and Caers (2010) used distance functions between pairs of patterns to quantify the characteristic statistics of a TI. Such metrics are easily incorporated in our inversion framework. We leave this for subsequent studies.

The use of the discrete cosine transform for model parameterization has the important advantage that the truncation level can be explicitly defined based on (a) the desired degree of small-scale structures that are to be resolved and (b) the sampling efficiency. The DREAM$_{(ZS)}$ algorithm is capable of handling a large number of parameters, but sampling efficiency will decrease with increasing dimensionality of the search space. The truncation levels of $n$ = 100 and $n$ = 150 used in our case studies are sufficient to adequately represent the subsurface structures. The truncation drastically reduces the dimensionality of the model space compared to a Cartesian representation, which requires the estimation of 5,000 and 11,200 cells in the channels and the fluvial deposits case, respectively. Despite their sparseness, the use of a TI-based parameterization ensures that dominant structures are preserved in the posterior models. Our present formulation is based on a user-defined value $n$ and it would be worthwhile to explore alternative formulations. Compressed sensing (e.g., Candès et al., 2006), in which the sparsity of the parameter coefficients are maximized, or transdimensional (reversible-jump) MCMC (e.g., Malinverno, 2002; Bodin and Sambridge, 2009), in which the number of model parameters are treated as an unknown throughout the inversion process, could be suitable starting points for developing a strategy to automatically estimate an appropriate value (or distribution) of $n$.

The main drawback of a DCT parameterization is that truncation impairs ability to accurately reproduce the sharp interfaces from the TI. Gradually updating a model according to an underlying geostatistical model (e.g., Mariethoz et al., 2010a; Hansen et al., 2012c) can help to preserve sharp interfaces, but such an approach requires geostatistical resimulation between model states and has to deal with strongly correlated posterior model realizations (Cordua et al., 2012; Ruggeri et al., 2013).

This study reveals some of the well-known problems of probabilistic inversion in high dimensional parameter spaces. Poorly estimated model parameters are easily compensated for by other parameters that are well defined. This we observe in the interplay between DCT coefficients and the petrophysical parameters. An important problem is that the parameter values used to generate the data are not necessarily the most likely ones (Figures 7 and 12). In other words, conditioning only on geophysical data (e.g. travel times) can produce misleading results. We show that incorporation of auxiliary information through the use of summary metrics from TI realizations significantly improves the posterior models and parameter values. The method can thus be seen as a strategy to handle high dimensionality in probabilistic inversion.

# 5   Conclusions

Training images are conceptual geological models of a certain site of interest. We use information from realizations of a TI to reduce the ambiguity inherent in probabilistic inversion of geophysical data by determining TI-based sparse model representations and by imposing morphological constraints on the posterior models. As a novelty, the model constraints are based on summarizing statistical metrics extracted from realizations of a TI. These are honored by formulating prior probability functions based on the distribution of summary metrics observed in the TI realizations. The summary metrics considered are global measures of model roughness, variability within the model and percentile distribution of porosity. Two different case studies were used to demonstrate the usefulness and applicability of the proposed inversion strategy. The use of summary statistics suppresses inversion artifacts, and provides parameter estimates that exhibit smaller posterior uncertainty and are in better agreement with their observed values. The methodology is computationally efficient, and designed to accommodate a wide range of geophysical forward problems and summary statistics. The summary metrics used in this study are rather simple, but significantly improved the reliability of the posterior parameter estimates and models. Application of more advanced summary metrics should hold even greater promise.

# Acknowledgements

This research was funded by the Swiss National Science Foundation (SNF)

and is a contribution to the ENSEMBLE project (grant no. CRSI22_132249). The second and third author acknowledge support from the UC-Lab Fees Research Program Award 237285. We thank Guillaume Pirot and Julien Straubhaar for sharing with us their geostatistical simulations. Detailed reviews from Malcolm Sambridge and an anonymous reviewer are greatly appreciated and helped to further refine the manuscript. This work is dedicated to the memory of Tobias Lochbühler who tragically lost his life in a mountaineering accident on July 19, 2014. Tobias was a truly wonderful person and a most gifted researcher.